# One-Dimensional Model for Coupled Flexural, Torsional and Extensional Motions of Elastic Beams with Embedded Point Magnets under Magnetic Fields

Jiashi Yang (jyang1@unl.edu)
Department of Mechanical and Materials Engineering
University of Nebraska-Lincoln, Lincoln, NE 68588-0526, USA

**Abstract**
This paper establishes a one-dimensional theoretical model for an elastic beam with embedded point magnets in extensional, bending and torsional motions. The beam has a circular cross-section. The point magnets may be in the interior or at the ends of the beam. They are assumed to be small and rigid, and may be ferromagnetic with spontaneous magnetic moments that can change their directions with respect to the magnets. The model shows that the extensional, flexural and torsional motions of the beam may be coupled by the point magnets when an external magnetic field is applied. A few examples are given. Piezoelectric beams with point magnets are also discussed.

## 1. Introduction

Elastic beams with discrete or point magnets at their ends or in the interior are common structures in various applications [1–3]. These structures often operate in an externally applied magnetic field which produces forces and couples on the magnets. Theoretical modeling of these structures presents a few challenges. The forces and couples on the magnets tend to cause coupled extensional, bending and torsional motions of the beam. Unless the beam has a circular cross-section, theoretical models for beams in coupled motions may be complicated because of the warping of beam cross-sections in torsion. When the magnets are saturated ferromagnets and the angular momentum of the magnetic moments need to be considered, complications arise when the saturated magnetic moments are allowed to change their directions with respect to the magnets. This paper is limited to the relatively simple case of coupled extensional, bending and torsional motions of an elastic beam with a circular cross-section and point magnets embedded in the interior or at the ends of the beam. The more complicated case of noncircular beams is left as future work. A zero-dimensional model is constructed for a small point magnet with consideration of its mass and moment of inertia as well as the angular momentum associated with the magnetic moment.

## 2. Zero-Dimensional Equations of a Small and Rigid Magnet

Consider a small and rigid ferromagnet with a volume $V$ and mass density $\rho$. Based on the two-continuum model in [4–10], it is assumed that the magnet consists of a rigid lattice continuum and a massless spin continuum. We denote the total mass of the magnet by $m$ and consider the magnetization vector $\mathbf{M}$ to be its average over $V$:

$$m = \int_V \rho(\mathbf{x})dV, \quad \mathbf{M}(t) = \frac{1}{V}\int_V \mathbf{M}(\mathbf{x},t)dV. \qquad (2.1)$$

The centroidal velocity of the magnet is denoted by $\mathbf{v}_G$ with respect to an inertial reference frame. We also introduce a centroidal reference frame for the magnet. The centroidal frame is in translation only without rotation. The rotations of the lattice continuum and the magnetic moment vector of the spin continuum in a motion are different in general, and are both assumed to be small. They are described by small angular displacements $\boldsymbol{\delta\theta}^l$ and $\boldsymbol{\delta\theta}^s$, respectively. The angular rates of the rotations of the lattice and the magnetic moment vector of the spin are denoted by

$$\boldsymbol{\omega}^l = \frac{\boldsymbol{\delta\theta}^l}{\delta t}, \quad \boldsymbol{\omega}^s = \frac{\boldsymbol{\delta\theta}^s}{\delta t}. \tag{2.2}$$

The moment of inertia per unit volume of the lattice with respect to its translational centroidal reference frame is denoted by a tensor $\mathbf{I}_G$. In general $\mathbf{I}_G$ is time dependent when the magnet is rotating with respect to its translational and centroidal frame. We assume a spherical magnet whose $(\mathbf{I}_G)_{ij} = I_G \delta_{ij}$ or $\mathbf{I}_G = I_G \mathbf{1}$ where $\mathbf{1}$ is the second-order unit tensor and $I_G$ is a constant. Then the angular momentum of the lattice is given by $\mathbf{I}_G \cdot \boldsymbol{\omega}^l$. The linear momentum equation for the magnet is

$$\frac{d}{dt}(m\mathbf{v}_G) = \Sigma \mathbf{F}, \tag{2.3}$$

where $\Sigma \mathbf{F}$ is the total force acting on the magnet. The angular momentum equations for the lattice and the spin separately are

$$\frac{d}{dt}(\mathbf{I}_G \cdot \boldsymbol{\omega}^l) = \boldsymbol{\Gamma}, \tag{2.4}$$

$$\frac{1}{\gamma}\frac{d\mathbf{M}}{dt} = \mathbf{M} \times \mathbf{B}^{eff}, \tag{2.5}$$

where $\boldsymbol{\Gamma}$ is the total couple on the lattice. $\mathbf{B}^{eff}$ is the total effective magnetic induction on the magnetization vector.

The interaction between the lattice and the spin is described by a local magnetic induction $\mathbf{B}^L$. The couple on the spin produced by $\mathbf{B}^L$ is $\mathbf{M} \times \mathbf{B}^L$. When unperturbed, the spontaneous magnetization $\mathbf{M}$ is along a direction of the lattice called an easy axis. Under a small disturbance, $\mathbf{M}$ deviates from the easy axis a little and experiences a restoring couple toward the easy axis. Effectively $\mathbf{M} \times \mathbf{B}^L$ plays the role of the restoring couple on the spin. For a simple description of the restoring couple we assume it is related to the relative rotation of the spin with respect to the lattice linearly, i.e.

$$\mathbf{M} \times \mathbf{B}^L = -\lambda M_s^2 (\boldsymbol{\delta\theta}^s - \boldsymbol{\delta\theta}^l) = -\lambda M_s^2 \boldsymbol{\delta\theta}^{s/l},$$
$$\boldsymbol{\delta\theta}^{s/l} = \boldsymbol{\delta\theta}^s - \boldsymbol{\delta\theta}^l, \tag{2.6}$$

where $\lambda$ is a material parameter and $M_s$ is the saturation magnetization. Multiplying both sides of Eq. (2.6) by $\mathbf{M}$ through a cross product, we have

$$(\mathbf{M} \times \mathbf{B}^L) \times \mathbf{M} = -\lambda M_s^2 \boldsymbol{\delta\theta}^{s/l} \times \mathbf{M}. \tag{2.7}$$

With the following vector identity:

$$\mathbf{A} \times (\mathbf{B} \times \mathbf{C}) = (\mathbf{A} \cdot \mathbf{C})\mathbf{B} - (\mathbf{A} \cdot \mathbf{B})\mathbf{C}, \tag{2.8}$$

we write Eq. (2.7) as

$$(\mathbf{M} \times \mathbf{B}^L) \times \mathbf{M} = -\mathbf{M} \times (\mathbf{M} \times \mathbf{B}^L) = \mathbf{M} \times (\mathbf{B}^L \times \mathbf{M})$$
$$= (\mathbf{M} \cdot \mathbf{M})\mathbf{B}^L - (\mathbf{M} \cdot \mathbf{B}^L)\mathbf{M} = M_s^2 \mathbf{B}^L - 0\mathbf{M} = M_s^2 \mathbf{B}^L = -\lambda M_s^2 \boldsymbol{\delta\theta}^{s/l} \times \mathbf{M}, \tag{2.9}$$

where $\mathbf{M} \cdot \mathbf{B}^L = 0$ has been used [4]. Hence

$$\mathbf{B}^L = -\lambda \boldsymbol{\delta\theta}^{s/l} \times \mathbf{M}, \tag{2.10}$$

which serves as the constitutive relation for $\mathbf{B}^L$ of the magnet. Equation (2.10) can be written as

$$\mathbf{B}^L = -\lambda(\delta\mathbf{M})^{s/l}, \qquad (2.11)$$

where

$$(\delta\mathbf{M})^{s/l} = \delta\boldsymbol{\theta}^{s/l} \times \mathbf{M}. \qquad (2.12)$$

When $\lambda = 0$, we have $\mathbf{B}^L = 0$. In this case there is no restoring couple. Any axis is an easy axis. When $\lambda = \infty$, from Eq. (2.6)$_1$, we have $\delta\boldsymbol{\theta}^{s/l} = 0$. In this case there is no relative rotation between the spin and the lattice or, in other words, the spin is fixed to the lattice. Using Eq. (2.11), we can write the restoring couple on $\mathbf{M}$ as

$$\mathbf{M} \times \mathbf{B}^L = -\lambda \mathbf{M} \times (\delta\mathbf{M})^{s/l}. \qquad (2.13)$$

## 3. One-Dimensional Equations for Elastic Beams

Consider the elastic beam in Fig. 1. It has a circular cross-section so that the one-dimensional theory for torsion in conventional mechanics of materials applies. The spatial coordinates are $x_k$ or $(x,y,z)$. The $x$ axis goes through the center of the circular cross-section. The material is assumed to be isotropic. We derive one-dimensional equations for extension, bending in the $(x,y)$ and $(x,z)$ planes as well as torsion in the following.

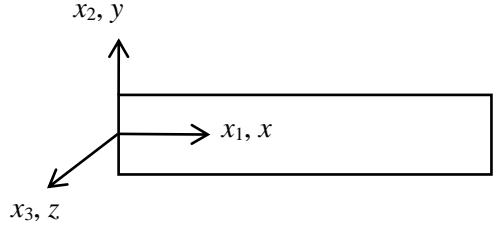

Fig. 1. An elastic beam with a circular cross-section.

For a one-dimensional model of the beam, the flexural displacements $u_2$ and $u_3$, axial displacement $u_1$ for extension and bending, and the angle of twist $\psi$ for torsion are approximated by [11,12]

$$\begin{aligned} u_2\big|_{y=z=0} &= v(x,t), \quad u_3\big|_{y=z=0} = w(x,t), \\ u_1(\mathbf{x},t) &= u(x,t) - yv' - zw', \\ \psi &= \psi(x,t). \end{aligned} \qquad (3.1)$$

The axial strain $\varepsilon$ and the shear strain $\gamma$ in torsion are related to $u$, $v$, $w$ and $\psi$ by

$$\varepsilon = u' - yv'' - zw'', \quad \gamma = r\psi', \qquad (3.2)$$

where a prime represents a differentiation with respect to $x$, the axial coordinate. $r$ and $\theta$ are polar coordinates within the circular cross-section. The axial stress $\sigma$ and the shear stress $\tau$ in torsion can be expressed in terms of $\varepsilon$ and $\gamma$ through stress-strain relations as

$$\begin{aligned} \sigma &= E\varepsilon = E(u' - yv'' - zw''), \\ \tau &= G\gamma = Gr\psi', \end{aligned} \qquad (3.3)$$

where $E$ is Young's modulus and $G$ the shear modulus. The internal forces and moments over a typical cross-section of the beam are shown in Fig. 2. In the figure, the moment vectors are represented by double arrows to distinguish them from force vectors.

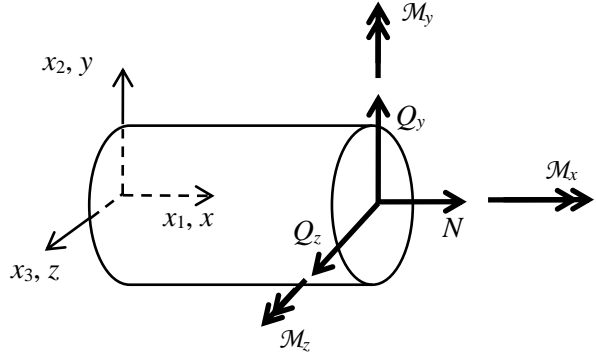

Fig. 2. Internal forces and moments over a cross-section.

The axial force $N$, twisting moment or torque $\mathcal{M}_x$ for torsion, and bending moments $\mathcal{M}_y$ and $\mathcal{M}_z$ are obtained by integrating the relevant stresses or their products with $y$, $z$ or $r$ over a cross-section [11,12]. They have the following expressions:

$$N = EAu', \quad \mathcal{M}_x = GI_p\psi', \quad \mathcal{M}_y = -EIw'', \quad \mathcal{M}_z = EIv'', \tag{3.4}$$

where $A$ is the area of the cross-section. $I$ and $I_p$ are the moment of inertia and polar moment of inertia of the cross-section. Let the radius of the circular cross-section be $a$. Then

$$A = \pi a^2, \quad I = \frac{\pi}{4}a^4, \quad I_p = \frac{\pi}{2}a^4. \tag{3.5}$$

The transverse shear forces $Q_y$ and $Q_z$ related to bending are determined from the following shear force-bending moment relations [11,12]:

$$Q_y = -\mathcal{M}_z' = -EIv''', \quad Q_z = \mathcal{M}_y' = -EIw'''. \tag{3.6}$$

The equations for extension, torsion, and deflections in the $y$ and $z$ directions are

$$N' + F_x = \rho A\ddot{u}, \quad \mathcal{M}_x' + m_x = \rho I_p\ddot{\psi}, \quad Q_y' + F_y = \rho A\ddot{v}, \quad Q_z' + F_z = \rho A\ddot{w}, \tag{3.7}$$

where $\rho$ is the mass density. $\mathbf{F}$ is distributed force per unit length of the beam. $m_x$ is distributed twisting moment per unit length. With substitutions from Eqs. (3.6) and (3.4), we can write (3.7) as four uncoupled equations for $u$, $\psi$, $v$ and $w$ and as:

$$EAu'' + F_x = \rho A\ddot{u}, \tag{3.8}$$

$$GI_p\psi'' + m_x = \rho I_p\ddot{\psi}, \tag{3.9}$$

$$-EIv'''' + F_y = \rho A\ddot{v}, \tag{3.10}$$

$$-EIw'''' + F_z = \rho A\ddot{w}, \tag{3.11}$$

where a homogeneous beam with uniform geometric and material properties has been assumed. The small rotations of a cross-section of the beam are denoted by

$$\boldsymbol{\beta} = \beta_x\mathbf{i} + \beta_y\mathbf{j} + \beta_z\mathbf{k}, \quad \beta_x = \psi, \quad \beta_y = -w', \quad \beta_z = v'. \tag{3.12}$$

Usual boundary conditions at the ends of a beam may be the prescriptions of

$$
\begin{aligned}
&u \quad \text{or} \quad N, \quad \psi \quad \text{or} \quad \mathcal{M}_x, \\
&v \quad \text{or} \quad Q_y, \quad \beta_z \quad \text{or} \quad \mathcal{M}_z, \\
&w \quad \text{or} \quad Q_z, \quad \beta_y \quad \text{or} \quad \mathcal{M}_y.
\end{aligned}
\quad (3.13)
$$

## 4. A Point Magnet in a Beam

Consider a point magnet at an interior point with $x = a$ of an elastic beam. We are interested in the motion of the magnet under an applied field $\mathbf{B}$. The field produced by the magnet is neglected. The free body diagram of the magnet is shown in Fig. 3. At the two cross-sections of the beam to the left and right of the magnet, there are extensional and shear forces as well as bending and twisting moments. The magnetic force $\mathbf{f}^M$ and couple $\mathbf{c}^M$ per unit volume of the magnet by an external magnetic induction $\mathbf{B}$ act on the spin continuum of the magnet. We have

$$\mathbf{f}^M = \mathbf{M} \cdot (\mathbf{B}\nabla), \quad \mathbf{c}^M = \mathbf{M} \times \mathbf{B}. \quad (4.1)$$

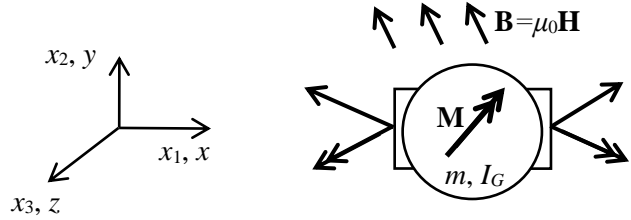

Fig. 3. Free body diagram of a magnet in a beam.

The linear and angular momentum equations of the magnet including both the lattice and spin continua are

$$
\begin{aligned}
N(a^+) - N(a^-) + f_x^M V &= m\ddot{u}, \\
Q_y(a^+) - Q_y(a^-) + f_y^M V &= m\ddot{v}, \\
Q_z(a^+) - Q_z(a^-) + f_z^M V &= m\ddot{w},
\end{aligned}
\quad (4.2)
$$

$$
\begin{aligned}
\mathcal{M}_x(a^+) - \mathcal{M}_x(a^-) + c_x^M V &= \dot{L}_x, \\
\mathcal{M}_y(a^+) - \mathcal{M}_y(a^-) + c_y^M V &= \dot{L}_y, \\
\mathcal{M}_z(a^+) - \mathcal{M}_z(a^-) + c_z^M V &= \dot{L}_z,
\end{aligned}
\quad (4.3)
$$

where $V$ is the volume of the magnet. $\mathbf{L}$ is the total angular momentum of the magnet including contributions from both the lattice and the spin. Equations (4.2) and (4.3) are the jump conditions across a point magnet at $x = a$.

The small rotation and angular velocity of the lattice of the magnet are the same as those of the beam cross-section in Eq. (3.12) at $x = a$, i.e.,

$$\delta\boldsymbol{\theta}^l = \boldsymbol{\beta}\big|_{x=a}, \quad \boldsymbol{\omega}^l = \dot{\boldsymbol{\beta}}\big|_{x=a}. \quad (4.4)$$

The total linear momentum of the magnet is from the lattice only. The total angular momentum of the magnet has contributions from both the lattice and the spin:

$$\mathbf{L} = I_G \boldsymbol{\omega}^l + \frac{1}{\gamma}\mathbf{M}V = I_G \dot{\boldsymbol{\beta}}\big|_{x=a} + \frac{1}{\gamma}\mathbf{M}V . \tag{4.5}$$

The time rate of change of **L** is given by

$$\dot{\mathbf{L}} = I_G \dot{\boldsymbol{\omega}}^l + \frac{1}{\gamma}\dot{\mathbf{M}}V = I_G \ddot{\boldsymbol{\beta}}\big|_{x=a} + \frac{1}{\gamma}\dot{\mathbf{M}}V . \tag{4.6}$$

$\dot{\mathbf{M}}$ is governed by the Landau–Lifshitz equation:

$$\frac{1}{\gamma}\dot{\mathbf{M}} = \mathbf{M} \times \mathbf{B}^{\text{eff}} = \mathbf{M} \times (\mathbf{B} + \mathbf{B}^L) , \tag{4.7}$$

where, from Eq. (2.11),

$$\mathbf{B}^L = -\lambda(\delta\mathbf{M})^{s/l} . \tag{4.8}$$

Under the motion of the beam and the applied magnetic induction **B** which is assumed to be small and known, the magnetization **M** differs from its finite reference state $\mathbf{M}^0$ by a small amount of $\mathbf{m} = \delta\mathbf{M}$:

$$\begin{aligned}\mathbf{M} &= \mathbf{M}^0 + \delta\mathbf{M} = \mathbf{M}^0 + \mathbf{m}, \\ \mathbf{m} &= \delta\mathbf{M} = \boldsymbol{\beta}\big|_{x=a} \times \mathbf{M}^0 + \delta\mathbf{M}^{s/l}.\end{aligned} \tag{4.9}$$

Since **m** is small, we have, approximately,

$$\mathbf{f}^M \cong \mathbf{M}^0 \cdot (\mathbf{B}\nabla), \quad \mathbf{c}^M \cong \mathbf{M}^0 \times \mathbf{B} , \tag{4.10}$$

which are known. The total angular momentum of the magnet may be written as

$$\mathbf{L} = I_G \dot{\boldsymbol{\beta}}\big|_{x=a} + \frac{1}{\gamma}(\mathbf{M}^0 + \mathbf{m})V . \tag{4.11}$$

The time rate of change of **L** is

$$\dot{\mathbf{L}} = I_G \ddot{\boldsymbol{\beta}}\big|_{x=a} + \frac{1}{\gamma}\dot{\mathbf{m}}V . \tag{4.12}$$

$\dot{\mathbf{m}}$ is governed by the Landau–Lifshitz equation in Eq. (4.7) which is approximated by:

$$\frac{1}{\gamma}\dot{\mathbf{m}} \cong \mathbf{M}^0 \times (\mathbf{B} + \mathbf{B}^L) , \tag{4.13}$$

where

$$\mathbf{B}^L = -\lambda(\delta\mathbf{M})^{s/l} = -\lambda\left(\mathbf{m} - \boldsymbol{\beta}\big|_{x=a} \times \mathbf{M}^0\right) . \tag{4.14}$$

Substituting Eq. (4.14) into Eq. (4.13), we obtain

$$\frac{1}{\gamma}\dot{\mathbf{m}} \cong \mathbf{M}^0 \times \left[\mathbf{B} - \lambda\left(\mathbf{m} - \boldsymbol{\beta}\big|_{x=a} \times \mathbf{M}^0\right)\right] . \tag{4.15}$$

In summary, the $\dot{\mathbf{L}}$ in the jump condition in Eq. (4.3) is given by Eq. (4.12) where **m** is governed by Eq. (4.15).

In the special case when the spin is fixed to the lattice, $\lambda = \infty$, $\delta\mathbf{M}^{s/l} = 0$, **m** has the following expression:

$$\mathbf{m} = \boldsymbol{\beta}\big|_{x=a} \times \mathbf{M}^0 , \tag{4.16}$$

and, from Eq. (4.12),

$$\dot{\mathbf{L}} = I_G \ddot{\boldsymbol{\beta}}\big|_{x=a} + \frac{1}{\gamma}\dot{\mathbf{m}}V = I_G \ddot{\boldsymbol{\beta}}\big|_{x=a} + \frac{1}{\gamma}\dot{\boldsymbol{\beta}}\big|_{x=a} \times \mathbf{M}^0 V . \tag{4.17}$$

## 5. Bending of an Elastic Beam with a Magnet

Consider the elastic cantilever in Fig. 4 with a fixed left end and a point magnet at its right end. The spin of the magnet is assumed to be fixed to its lattice. The beam and the initial $\mathbf{M}^0$ define a plane. The plane containing the beam and $\mathbf{M}^0$ is chosen to be the $(x,y)$ plane. The applied $\mathbf{B}$ is assumed to be in the $(x,y)$ plane too. $\mathbf{B}$ is also assumed to be static and uniform. The effect of $\mathbf{M}^0$ on $\mathbf{B}$ is neglected. In this case $\mathbf{M}^0$ experiences a couple only, i.e.,

$$\mathbf{f}^M = \mathbf{M} \cdot (\mathbf{B}\nabla) = 0,$$
$$\mathbf{c}^M = \mathbf{M} \times \mathbf{B} \cong \mathbf{M}^0 \times \mathbf{B} = (M_x^0 \mathbf{i} + M_y^0 \mathbf{j}) \times (B_x \mathbf{i} + B_y \mathbf{j}) = (M_x^0 B_y - M_y^0 B_x)\mathbf{k} = c_z^M \mathbf{k}. \quad (5.1)$$

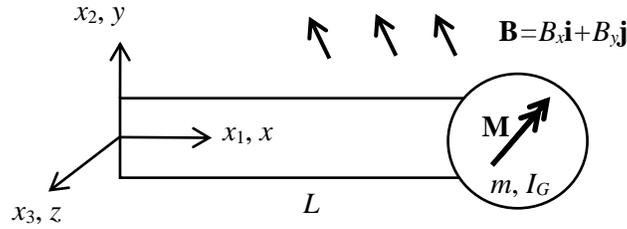

Fig. 4. An elastic cantilever with an end magnet in a magnetic field in the $(x,y)$ plane.

In this case the end magnet is equivalent to an end bending moment. Hence the beam is in bending in the $(x,y)$ plane with a deflection curve described by $v(x)$. For static bending, the boundary-value problem is

$$-EIv'''' = 0, \quad 0 < x < L,$$
$$v(0) = 0,$$
$$\beta_z(0) = v'(0) = 0, \quad (5.2)$$
$$Q_y(L) = -EIv'''(L) = 0,$$
$$-\mathcal{M}_z(L^-) + c_z^M V = -EIv''(L^-) + (M_x^0 B_y - M_y^0 B_x)V = 0.$$

The beam is in pure bending with a constant curvature. The solution for the deflection curve from Eq. (5.2) is

$$v(x) = \frac{x^2}{2EI}(M_x^0 B_y - M_y^0 B_x)V. \quad (5.3)$$

## 6. Torsion of an Elastic Beam with a Magnet

Consider the elastic cantilever with an end magnet in Fig. 5. The initial $\mathbf{M}^0 = M_y^0 \mathbf{j}$. The applied magnetic induction is $\mathbf{B} = B_z \mathbf{k}$ which is static and uniform.

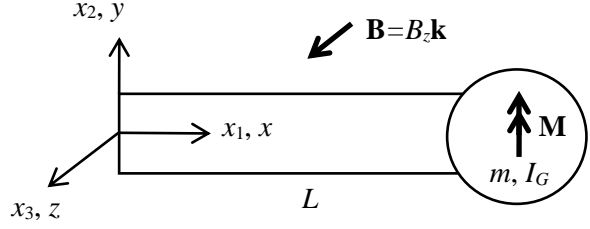

Fig. 5. An elastic cantilever with an end magnet in a magnetic field along the $z$ axis.

The magnetic loads on $\mathbf{M}^0$ are given by
$$\mathbf{f}^M = \mathbf{M} \cdot (\mathbf{B}\nabla) = 0,$$
$$\mathbf{c}^M = \mathbf{M} \times \mathbf{B} \cong \mathbf{M}^0 \times \mathbf{B} = M_y^0 \mathbf{j} \times B_z \mathbf{k} = M_y^0 B_z \mathbf{i} = c_x^M \mathbf{i}. \qquad (6.1)$$

In this case the end magnet is equivalent to an end twisting moment. The beam is in static torsion which is governed by
$$GI_p \psi'' = 0, \quad 0 < x < L,$$
$$\beta_x(0) = \psi(0) = 0, \qquad (6.2)$$
$$-\mathcal{M}_x(L^-) + c_x^M V = -GI_p \psi'(L) + M_y^0 B_z V = 0.$$

The solution for the angle of twist is
$$\psi(x) = \frac{M_y^0 B_z V}{GI_p} x. \qquad (6.3)$$

### 7. Elastically Connected Magnets

Elastically interacting magnets are common structures in devices [2,3]. Figure 6 shows a simple example. The magnets are sufficiently far away from each other so that their magnetic interactions are neglected. The applied $\mathbf{B}$ is static and uniform.

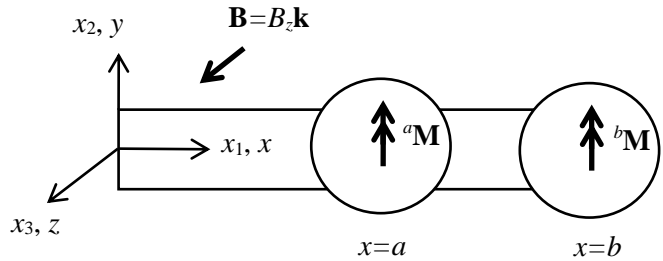

Fig. 6. A beam with two magnets in torsion.

The basic approach of analyzing a beam with an interior concentrated load is to break the beam into two parts at $x = a$, solve the governing equations in each part, impose boundary conditions at both ends, and apply continuity/jump conditions at $x = a$. Since the problem is linear, we use the method of superposition based on the solution of the uniform torsion problem in the previous section. The twisting moments at $x = a$ and $b$ are

$$^a\mathbf{c}^M \cong {^aM_y^0}\mathbf{j} \times B_z\mathbf{k} = {^aM_y^0}B_z\mathbf{i} = {^ac_x^M}\mathbf{i},$$
$$^b\mathbf{c}^M \cong {^bM_y^0}\mathbf{j} \times B_z\mathbf{k} = {^bM_y^0}B_z\mathbf{i} = {^bc_x^M}\mathbf{i}.$$
(7.1)

When the two magnets exist separately, we have

$$\psi(x, {^a\mathbf{M}}) = \begin{cases} \dfrac{{^aM_y^0} B_z {^aV}}{GI_p} x, & 0 < x < a, \\ \dfrac{{^aM_y^0} B_z {^aV}}{GI_p} a, & a < x < b, \end{cases}$$
(7.2)

$$\psi(x, {^b\mathbf{M}}) = \dfrac{{^bM_y^0} B_z {^bV}}{GI_p} x, \quad 0 < x < b.$$
(7.3)

When the two magnets are both present, we add Eqs. (7.2) and (7.3) to obtain

$$\psi(x, {^a\mathbf{M}}, {^b\mathbf{M}}) = \psi(x, {^a\mathbf{M}}) + \psi(x, {^b\mathbf{M}}) = \begin{cases} \dfrac{{^aM_y^0} B_z {^aV}}{GI_p} x + \dfrac{{^bM_y^0} B_z {^bV}}{GI_p} x, & 0 < x < a, \\ \dfrac{{^aM_y^0} B_z {^aV}}{GI_p} a + \dfrac{{^bM_y^0} B_z {^bV}}{GI_p} x, & a < x < b. \end{cases}$$
(7.4)

The rotation at $x = b$ due to $^a\mathbf{M}$ alone is

$$\psi(b, {^a\mathbf{M}}, 0) = \dfrac{{^aM_y^0} B_z {^aV}}{GI_p} a.$$
(7.5)

The rotation at $x = a$ due to $^b\mathbf{M}$ alone is

$$\psi(a, 0, {^b\mathbf{M}}) = \dfrac{{^bM_y^0} B_z {^bV}}{GI_p} a.$$
(7.6)

We note that

$$\dfrac{\psi(b, {^a\mathbf{M}}, 0)}{{^aM_y^0} B_z {^aV}} = \dfrac{a}{GI_p} = \dfrac{\psi(a, 0, {^b\mathbf{M}})}{{^bM_y^0} B_z {^bV}},$$
(7.7)

which is known as reciprocity.

## 8. Bending of a Piezoelectric Beam with a Magnet

Consider the piezoelectric beam in Fig. 7. It has two layers of polarized ceramics with opposite poling directions and is called a bimorph [13,14]. When the poling direction is reversed, the elastic and dielectric constants remain the same but the piezoelectric constants change their signs. Under a voltage $\mathcal{V}(t)$ across the electrodes at the top and bottom of the bimorph, one piezoelectric layer elongates and the other contracts along $x$ or vice versa. Hence bending in the $(x,z)$ plane is produced piezoelectrically. The beam has an end magnet whose $\mathbf{M}^0$ and the applied $\mathbf{B}$ are both in the $(x,z)$ plane. We consider the case when $\mathcal{V}$ and $\mathbf{B}$ are static and $\mathbf{B}$ is uniform.

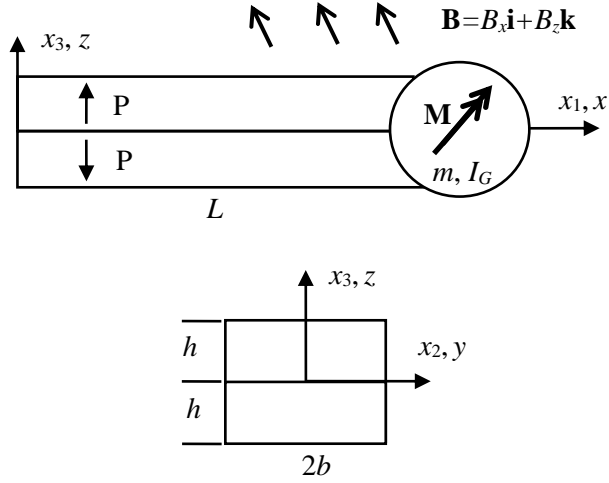

Fig. 7. A piezoelectric cantilever and its cross-section.

The deflection of the beam is described by
$$u_3 \cong u_3(x_1, t). \tag{8.1}$$
The bending moment $\mathcal{M} = \mathcal{M}_y$, shear force $Q = Q_z$ and the transvers load $F_3$ per unit length of the beam are shown in Fig. 8. The expressions of $\mathcal{M}$ and $Q$ are [13,14]
$$\mathcal{M} = \int_A T_{11} x_3 dA = -\bar{c}_{11} I u_{3,11} + \bar{e}_{31} bh\mathcal{V}, \tag{8.2}$$
$$Q = \frac{\partial \mathcal{M}}{\partial x_1} = -\bar{c}_{11} I u_{3,111}, \tag{8.3}$$
where $\bar{c}_{11}$ and $\bar{e}_{31}$ are the effective one-dimensional elastic and piezoelectric constants:
$$\bar{c}_{11} = \frac{1}{s_{11}}, \quad \bar{e}_{31} = \frac{d_{31}}{s_{11}}, \quad A = 4bh, \quad I = \frac{4bh^3}{3}. \tag{8.4}$$
$d_{31}$ is the relevant three-dimensional piezoelectric constant in matrix notation [13,14].

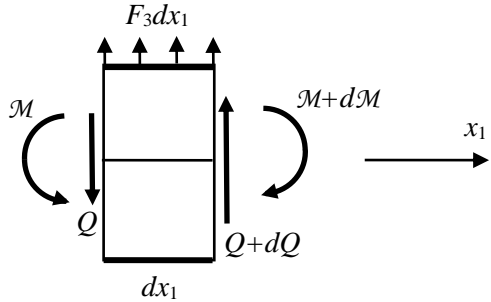

Fig. 8. A differential element of the beam and mechanical loads.

The equation for the flexural motion of the beam is obtained by applying Newton's second law to the differential element in Fig. 8 in the $x_3$ direction:

$$\frac{\partial Q}{\partial x_1} + F_3 = \rho A \ddot{u}_3. \tag{8.5}$$

The substitution of Eq. (8.3) into Eq. (8.5) yields an equation for $u_3$:

$$-I\bar{c}_{11}u_{3,1111} + F_3 = \rho A \ddot{u}_3. \tag{8.6}$$

The magnetic loads on $\mathbf{M}^0$ are given by

$$\begin{aligned}\mathbf{f}^M &= \mathbf{M} \cdot (\mathbf{B}\nabla) = 0, \\ \mathbf{c}^M &= \mathbf{M} \times \mathbf{B} \cong \mathbf{M}^0 \times \mathbf{B} = (M_x^0 \mathbf{i} + M_z^0 \mathbf{k}) \times (B_x \mathbf{i} + B_z \mathbf{k}) = (M_z^0 B_x - M_x^0 B_z)\mathbf{j} = c_y^M \mathbf{j}.\end{aligned} \tag{8.7}$$

The boundary-value problem for static bending under $\mathcal{V}$ and $\mathbf{B}$ is

$$-\bar{c}_{11}Iu_{3,1111} = 0, \quad 0 < x < L, \tag{8.8}$$

$$u_3(0) = 0,$$
$$\beta_y(0) = -u_{3,1}(0) = 0, \tag{8.9}$$

$$Q_z(L) = -\bar{c}_{11}Iu_{3,111}(L) = 0,$$
$$-\mathcal{M}_y(L^-) + c_y^M V = \bar{c}_{11}Iu_{3,11}(L^-) - \bar{e}_{31}bh\mathcal{V} + (M_z^0 B_x - M_x^0 B_z)V = 0, \tag{8.10}$$

where Eq. (4.3)$_2$ has been used. It can be seen that $\mathcal{V}$ and $\mathbf{B}$ produce an effective bending moment at the right end. The solution for the deflection curve is

$$u_3(x) = \frac{x^2}{2\bar{c}_{11}I}\left[\bar{e}_{31}bh\mathcal{V} - (M_z^0 B_x - M_x^0 B_z)V\right]. \tag{8.11}$$

## 9. Torsion of a Piezoelectric Beam with a Magnet

Consider the circular cylindrical beam of inner radius $a$ and outer radius $b$ shown in Fig. 9. The cross-section is denoted by $A$. The cylinder is made of polarized ceramics with circumferential poling along $\theta$. We choose $(r,\theta,z)$ to correspond to indices (2,3,1) so that the poling direction corresponds to 3. The lateral cylindrical surfaces of the beam are traction free and are unelectroded. The electric field in the free-space is neglected as an approximation. $\mathbf{B}$ is static and uniform.

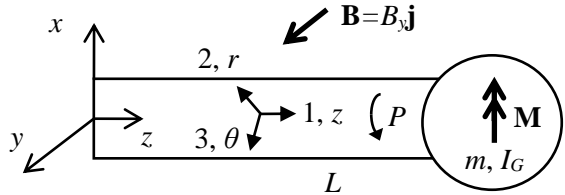

Fig. 9. A piezoelectric cantilever with an end magnet in a magnetic field along the $y$ axis.

The magnetic loads on $\mathbf{M}^0$ are given by

$$\begin{aligned}\mathbf{f}^M &= \mathbf{M} \cdot (\mathbf{B}\nabla) = 0, \\ \mathbf{c}^M &= \mathbf{M} \times \mathbf{B} \cong \mathbf{M}^0 \times \mathbf{B} = M_x^0 \mathbf{i} \times B_y \mathbf{j} = M_x^0 B_y \mathbf{k} = c_z^M \mathbf{k}.\end{aligned} \tag{9.1}$$

The angle of twist and the electric potential are approximated by functions of the axial coordinate $z$ and time only [13,14]:

$$\psi = \psi(z,t), \quad \varphi = \varphi(z,t). \tag{9.2}$$

The twisting moment $\mathcal{M} = \mathcal{M}_z$ and the axial electric displacement $\hat{D}_z$ over a cross-section are given by [13,14]:

$$\mathcal{M} = \mathcal{M}_z = \int_A T_{z\theta} r dA = c_{55} I_p \psi_{,z} + e_{15} C \varphi_{,z}, \tag{9.3}$$

$$\hat{D}_z = \int_A D_z dA = e_{15} C \psi_{,z} - \varepsilon_{11} A \varphi_{,z}, \tag{9.4}$$

where $c_{55}$ is the relevant shear elastic constant, $e_{15}$ the relevant piezoelectric constant, $\varepsilon_{11}$ the relevant dielectric constant, and

$$A = \pi(b^2 - a^2), \quad C = \frac{2\pi}{3}(b^3 - a^3), \quad I_p = \frac{\pi}{2}(b^4 - a^4). \tag{9.5}$$

The equation for torsional motion is obtained by the moment equation of the differential element in Fig. 10:

$$\mathcal{M}_{,z} + m_z = \rho I_p \ddot{\psi}, \tag{9.6}$$

where $m_z(z,t)$ is the distributed torsional load per unit length of the beam.

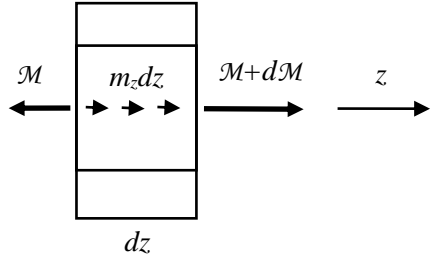

Fig. 10. A differential element of the beam under torsional loads.

Similarly, the one-dimensional charge equation of electrostatics can be obtained by applying electric loads to the differential element in Fig. 10 [13,14]:

$$\hat{D}_{z,z} = 0. \tag{9.7}$$

Substituting Eqs. (9.3) and (9.4) into Eqs. (9.6) and (9.7), for a uniform beam, we arrive at two equations for $\psi$ and $\varphi$ as

$$\begin{aligned} c_{55} I_p \psi_{,zz} + e_{15} C \varphi_{,zz} + m_z &= \rho I_p \ddot{\psi}, \\ e_{15} C \psi_{,zz} - \varepsilon_{11} A \varphi_{,zz} &= 0. \end{aligned} \tag{9.8}$$

For static torsion under **B** and a constant voltage $\mathcal{V}$ across the two ends of the beam, the boundary-value problem is

$$\begin{aligned} c_{55} I_p \psi_{,zz} + e_{15} C \varphi_{,zz} &= 0, \quad 0 < z < L, \\ e_{15} C \psi_{,zz} - \varepsilon_{11} A \varphi_{,zz} &= 0, \quad 0 < z < L. \end{aligned} \tag{9.9}$$

$$\begin{aligned} \varphi(0) &= 0, \\ \beta_z(0) &= \psi(0) = 0, \end{aligned} \tag{9.10}$$

$$\begin{aligned} \varphi(L) &= \mathcal{V}, \\ -\mathcal{M}_z(L^-) + c_z^M V &= -[c_{55} I_p \psi_{,z}(L) + e_{15} C \varphi_{,z}(L)] + M_x^0 B_y V = 0, \end{aligned} \tag{9.11}$$

where the left end is fixed mechanically and grounded electrically, and Eq. (4.3)$_3$ has been used. It can be seen that $\mathcal{V}$ and **B** produce an effective twisting moment at the right end. The solution for the electric potential $\varphi$ and the angle of twist $\psi$ is

$$\varphi(x) = \frac{\mathcal{V}}{L} z,$$
$$\psi(x) = \frac{z}{c_{55} I_p} \left[ -e_{15} C \frac{\mathcal{V}}{L} + M_x^0 B_y V \right]. \qquad (9.12)$$

## 10. Conclusions

A point magnet in an applied magnetic field experiences a force and a couple. In the case of a beam with point magnetics in a uniform magnetic field, the forces on the magnets vanish and the couples remain which tend to cause coupling between torsion and bending of the beam. For the beam with a circular-cross section treated in this paper, coupled torsional and flexural motions are relatively simple. For a beam with a noncircular cross-section, a more sophisticated model is needed because of the warping of the cross-section of the beam associated with torsion. Similar to one-dimensional models for beams with point magnets, two-dimensional models of thin films or plates with point magnets are also needed because of their applications.